\documentclass[12pt]{article}
\usepackage[utf8]{inputenc}
\usepackage[T1]{fontenc}
\usepackage[margin=1in]{geometry}
\usepackage{caption}
\usepackage{subcaption}
\usepackage{graphicx}
\usepackage{xcolor}
\usepackage{amsmath}
\usepackage{amssymb}
\usepackage{booktabs}
 
\begin{document}
 
\begin{center}
    {\LARGE Comparison of melanin subtypes using pump-probe microscopy\par}
    \vspace{1.5em}
    {\large Emma Bao\textsuperscript{1,*}, David Grass\textsuperscript{2},
    Martin C. Fischer\textsuperscript{1,2}, and Warren S. Warren\textsuperscript{1,2,3,4}\par}
    \vspace{1em}
    {\small
    \textsuperscript{1}Department of Physics, Duke University, Durham, North Carolina, USA\\
    \textsuperscript{2}Department of Chemistry, Duke University, Durham, North Carolina, USA\\
    \textsuperscript{3}Department of Biomedical Engineering, Duke University, Durham, North Carolina, USA\\
    \textsuperscript{4}Department of Radiology, Duke University, Durham, North Carolina, USA\\[1ex]
    \textsuperscript{*}emma.bao@duke.edu\par}
\end{center}

\begin{abstract} 
Pump-probe microscopy (PPM) is used to probe excited state dynamics of endogenous chromophores to generate contrast that is unavailable in traditional forms of microscopy. Here we apply PPM to characterize melanin, a pigment central to photoprotection, skin and eye pigmentation, and melanoma pathogenesis. Melanin is notoriously difficult to analyze due to its structural and chemical complexity. We imaged melanin from six different lesion categories: cutaneous nevi, lentigo, cutaneous malignant melanoma, conjunctival melanocytic lesions, vulvar melanocytic lesions, and a murine melanoma model, across 21 independent biological samples and 109 regions of interest. Using model fitting, adapted phasor analysis, and support vector machine classification, we characterize and quantify excited-state dynamics as a surrogate for chemical and structural heterogeneity. Heterogeneity of melanin separates cutaneous from non-cutaneous lesions, but is not correlated with malignancy. Benign lentigo is the most heterogeneous population in the dataset, exceeding cutaneous malignant melanoma, while conjunctival lesions are the most homogeneous despite receiving direct solar exposure. Classification on full transient absorption curves separates most subtypes well above the six-class chance level. These findings demonstrate the ability of PPM to reveal subtle differences in melanin photophysics across diverse tissue environments, with implications for optical tissue classification and the understanding of melanoma pathogenesis.
\end{abstract}

\section{Introduction}
Melanin is a natural pigment that is responsible for coloration of skin, hair, and eyes, photoprotection from ultraviolet radiation, and binding of heavy metals\cite{cichorek2013skin, brenner2008protective, hong_current_2007, sarna_interaction_2022}. Most melanin is synthesized through the oxidation and polymerization of tyrosine and phenol derivatives in a process known as melanogenesis within specialized cells called melanocytes \cite{cichorek2013skin}. In humans, melanin occurs primarily as two chemically distinct pigments: eumelanin, a black-to-brown pigment derived from the oxidative polymerization of L-dopa, and pheomelanin, a yellow-to-reddish brown pigment derived from oxidation of cysteinyl-dopa precursors. During eumelanogenesis, dopachrome is a key intermediate that undergoes decarboxylation to form 5,6-dihydroxyindole (DHI) or tautomerization to form 5,6-dihydroxyindole-2-carboxylic acid (DHICA)\cite{d2013melanins}. DHI and DHICA oxidatively couple into small oligomers, which oxidize further into nearly planar sheets\cite{Ju2019}. Four to five of these sheets, each containing four to eight indole units, stack with a graphite-like interlayer spacing of 3.4 \AA\ to form protomolecules approximately 15 \AA\ in size \cite{d2009chemical}. Melanocytes package melanin into melanosomes, small organelles of approximately $1$ $\mu$m size, and deliver them to neighboring epithelial cells, including keratinocytes in cutaneous and vulvar tissue and non-keratinized conjunctival epithelial cells in conjunctival tissue \cite{bento2023melanin, whittington2024melanocytic}. 

Melanocytes share a common neural crest origin, yet their differentiation is also shaped by anatomical location and local microenvironment, giving rise to biologically distinct melanocytic populations \cite{cichorek2013skin}. In the skin, ultraviolet radiation (UVR) exposure drives increased melanin production as part of the photoprotective response, but it also introduces heterogeneity in both pigment distribution and the underlying molecular structure of melanin \cite{brenner2008protective}. 
Conjunctival melanocytes lie at the conjunctival surface, and transfer melanosomes to the overlying non-keratinized epithelium. Unlike uveal and retinal pigments, they receive direct solar exposure \cite{pandiani2017focus}. However, conjunctival melanomas display a unique tissue architecture not found in the skin. The conjunctiva consists of a stratified non-keratinized epithelium lacking rete ridges where the concepts of papillary and reticular stromal organization are not applicable \cite{whittington2024melanocytic}. Although UVR has been implicated indirectly in the pathogenesis of vulvar melanoma, potentially affecting local immune responses, its contribution is far less significant than in cutaneous melanoma because vulvar tissues are generally not exposed to UVR \cite{dobricua2021vulvar, lorenz2022cutaneous, wohlmuth2021vulvar}. Prior pump–probe studies have shown that the pump‑probe signatures, interpreted as differences in melanin composition and aggregation, differ between melanomas that metastasize and those that do not, underscoring the biological relevance of melanin heterogeneity \cite{robles2015pump, Ju2019, grass2022contrast}.

Due to the broad and featureless absorption spectrum of melanin, most conventional imaging techniques are not well suited to study melanin, particularly its spatial heterogeneity. Linear techniques such as reflectance confocal microscopy and optical coherence tomography lack chemical specificity \cite{birmpili2025optical}. Multiphoton fluorescence lifetime imaging and Raman spectroscopy have been used to analyze cellular and biochemical changes in melanin exposed to UVR \cite{pena_multiphoton_2023, wilkinson2022fingerprinting, ali2013raman}. Fluorescence lifetime imaging microscopy, harmonic generation, and coherent Raman scattering have been successfully applied to melanocytic samples \cite{zhou2025recent,Liao2019,Shirshin2022, sitiwin_shedding_2019, pena_vivo_2022}, but analysis of the chemical heterogeneity across different melanin populations has not yet been reported, to the best of our knowledge. A comprehensive review about melanin characterization was recently published by Ito and Wakamatsu \cite{wakamatsu_recent_2023}. Pump-probe microscopy (PPM) has been used to distinguish melanomas that metastasize from melanomas that do not \cite{robles2015pump, matthews2011pump, Ju2019}, to investigate the chemical structure and building blocks of melanin \cite{Simpson2013, Simpson2014, Ju2018, grass2022contrast}, and to image melanin in specific lesions such as conjunctival and vulvar melanoma \cite{wilson_imaging_2013, Robles2017}. These studies addressed either diagnosis within a single lesion type or synthetic melanin standards, leaving the broader comparison of biologically distinct melanin populations unexplored. In this manuscript, we use PPM to systematically characterize melanin's chemical heterogeneity across a diverse set of melanocytic lesions, including conjunctival melanocytic lesions, vulvar melanoma, nevi, cutaneous malignant melanoma, lentigo, and a mouse melanoma model. These specimens represent biologically and anatomically distinct melanin populations.

\section{Materials and Methods}
    \subsection{Pump-Probe Microscopy}
    \begin{figure}[h]
        \centering
        \includegraphics[width = \linewidth]{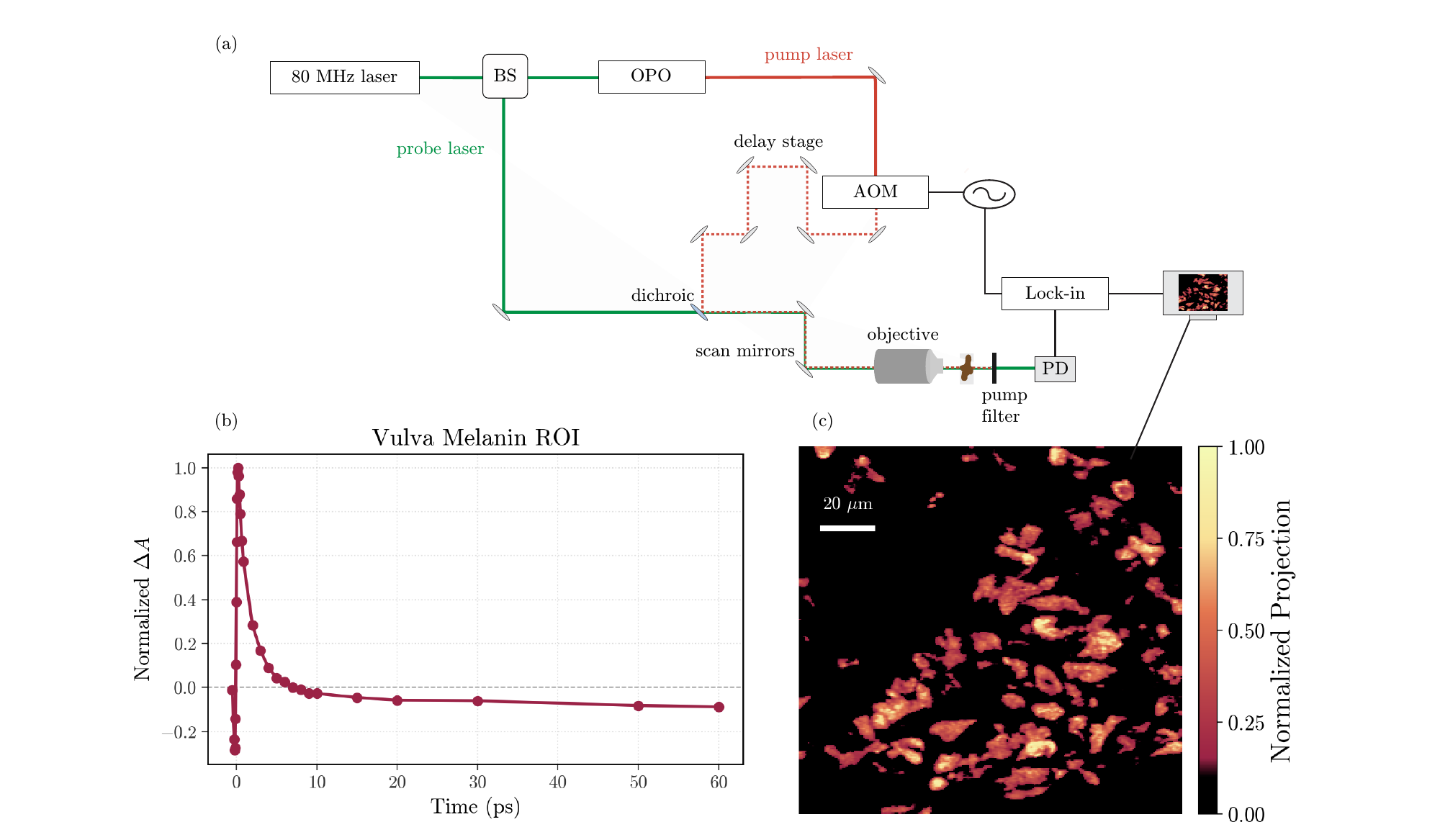}
        \caption{(a) Experimental pump-probe microscope setup. (b) Spatially averaged transient absorption (TA) curve of a vulvar melanoma biopsy. (c) Normalized intensity projection of PPM image stack of a vulvar melanoma biopsy.}
        \label{fig:setup}
    \end{figure}
    Pump-probe microscopy (PPM) is a nonlinear optical technique that probes excited-state dynamics of molecules with subcellular spatial resolution and sub-picosecond temporal resolution \cite{fischer2016invited}. The technique measures transient absorption: a pump pulse generates excited electronic or vibrational populations in the sample, thereby changing the absorption cross-section at the probe wavelength. By recording the probe transmission as a function of time delay $\Delta t$ between pump and probe at each image pixel, one obtains a transient absorption (or pump-probe) image stack. These dynamics depend on molecular structure, aggregation state, and chemical environment, and transient absorption curves can provide chemically specific contrast inaccessible to linear optical techniques. 
    
    Our PPM implementation is based on a tunable, mode-locked Ti:Sapphire laser (Coherent Chameleon) producing a pulse train at a repetition rate of 80~MHz. The output is split into two beam paths: a probe beam with a wavelength $\lambda_\text{probe}$ and, after passing an optical parametric oscillator (Coherent MIRA-OPO), a tunable pump beam $\lambda_\text{pump}$ as shown in figure~\ref{fig:setup}. The pump pulse train is intensity-modulated at 2~MHz by an acousto-optic modulator (AOM), and the interpulse delay $\Delta t$ between pump and probe is set by a motorized delay stage. Both beams are spatially overlapped at a beam splitter and directed into a laser scanning microscope, where a 20$\times$ objective (NA = 0.7, Olympus) focuses them onto the sample. For each fixed delay $\Delta t$, the beams are raster-scanned over the sample, and the procedure is repeated across all delays to yield a spatiotemporal image stack per region of interest. After rejecting the pump beam with a dichroic filter, the modulated probe is detected by an amplified photodiode and demodulated by a lock-in amplifier referenced to the AOM frequency, recovering the change in probe transmission at each pixel. Transient absorption is detected via modulation transfer \cite{fischer2016invited}: the nonlinear interaction between pump and probe at the sample transfers some of the 2~MHz pump modulation onto the probe intensity. Figure~\ref{fig:setup}b-c display a typical PPM image stack: (b) shows the spatially averaged PPM stack highlighting typical melanin transients and (c) represents the intensity projection over all time delays.
    
    \subsection{Samples and Data Acquisition}
    Pump-probe images were acquired using a pump wavelength of $\lambda_\text{pump}=720$ nm and probe wavelength of $\lambda_\text{probe}=817$ nm, each at powers of $0.3$ mW with a beam waist of $0.5 \text{ }\mu$m with parallel polarization. This wavelength pair has previously demonstrated sensitivity to the broad absorption spectrum of melanin and its aggregation state \cite{Ju2019, Simpson2014, matthews2011pump, robles2013quantifying}. Each acquisition consisted of a raster scan over a $155 \mu\text{m}\times  155 \mu\text{m}$ field of view at 36 interpulse delays $\Delta t$ spanning $-15$ to $60$ ps, yielding a spatiotemporal image stack.
    
    A total of 109 regions of interest (ROIs) were acquired from melanin of six lesion categories: cutaneous melanoma (n = 5), nevi (n = 4), lentigo (n = 3), conjunctival melanoma (n = 3), vulvar melanoma (n = 3), and murine melanoma (n = 3). They were selected to represent the chemical and structural diversity of melanin across tissue compartments and disease states. Melanin in cutaneous tissue and melanin in conjunctival tissues differ substantially in their eumelanin-to-pheomelanin ratio, their DHI:DHICA composition (the two indole monomers of eumelanin), and their metal chelation state \cite{Ito2003,hong_current_2007}, while lentigo and vulvar melanin represent less-studied pigmented phenotypes. The murine samples were derived from a BRAF V600E-inducible transgenic model in which melanocyte-specific oncogene expression was triggered by topical 4-hydroxytamoxifen (4-OHT) application \cite{goel2009melanocytic}. Conjunctival melanocytic lesions arise from melanocytes within the conjunctival mucosa and exhibit unique molecular drivers and histologic features compared with their cutaneous counterparts \cite{whittington2024melanocytic, wilson_imaging_2013}. Vulvar melanocytic lesions originate from melanocytes of the vulvovaginal mucosa, which reside in a microenvironment that differs from the skin in terms of immune regulation, environmental exposure, and molecular characteristics \cite{Robles2017, wohlmuth2021vulvar, hassan2025mucosal}. Lentigines are characterized by increased melanocyte activity and a localized buildup of melanin within the basal epidermis caused by chronic UV exposure that enhances melanogenesis, resulting in denser and more irregular pigment deposition than in surrounding skin \cite{cardinali2012mechanisms, mardani2025treatment, mizutani2021three}.
    
    All human specimens were formalin-fixed and paraffin-embedded (FFPE) following standard histopathological protocols. Briefly, tissues were fixed in 10\% neutral-buffered formalin, dehydrated, and embedded in paraffin. Two adjacent 5 $\mu$m sections were cut from each FFPE block: one was stained with hematoxylin and eosin (H\&E) for pathological review, and the adjacent section was deparaffinized and mounted between a glass slide and coverslip without staining for pump-probe imaging. A board-certified pathologist reviewed the H\&E sections and annotated tumor regions, which were then co-registered to the unstained adjacent section to define ROIs for imaging. 
    
    \subsection{Data Processing}
    Each pump-probe image stack undergoes preprocessing including background subtraction, thresholding, and binning prior to three complementary analyses (see Supplementary material): exponential model fitting, adapted phasor analysis, and support vector machine classification.
    \subsubsection{Fitting}
    \label{ch:fitting}
    Each of the 109 image stacks is spatially averaged across all pixels within each ROI to yield a single representative transient absorption (TA) curve. The TA curve is fit to a parametric model to extract characteristic lifetimes of the underlying excited-state dynamics. The fit function $f(\Delta t)$ is selected by the Akaike information criterion \cite{Akaike1974, Cavanaugh1997} and is composed of a superposition of an instantaneous component with a lifetime much shorter than the temporal resolution of the microscope, modeled as a delta function $a_0 \delta(t)$, three exponential decays $a_i e^{-t/ t_i}$ with $i\in\{1, 2, 3\}$, and a long-lived offset component $a_\infty$, corresponding to an exponential decay whose lifetime far exceeds the acquisition window. The exponential and offset terms are multiplied by the Heaviside step function $\theta(t)$ so that the response begins at zero delay. The full model is convolved with the instrument response function (IRF) to account for the finite pulse widths of the pump and probe beams and delays in the data acquisition system:
    \begin{align}
  f(\Delta t) = \left[ \mathrm{IRF} \ast \left( a_0\,\delta(t) + \theta(t)\left[ \sum_{i=1}^{3} a_i\, e^{-t/t_i} + a_\infty \right] \right) \right](\Delta t). 
    \end{align}
    The fit procedure follows \cite{grass2022contrast}; a detailed description is provided in Supplementary material. The extracted characteristic lifetimes are sorted and predominantly fall into three classes: $t_1$ ($< 1$ ps), $t_2$ ($1-5$ ps), and $t_3$ ($> 5$ ps).  For a subset of image stacks, an interplay between the slowest finite lifetime $t_3$ and the long-lived offset component produced fitting instabilities. In these cases, a reduced model with two exponential decays was used instead.

    \subsubsection{Adapted Phasor Analysis}
    \label{ch:phasor}
    Adapted phasor analysis\cite{robles2012phasor}, originally developed for fluorescence lifetime imaging microscopy \cite{digman_phasor_2008}, offers a model-free approach to visualizing and quantifying complex multiexponential dynamics. The time-resolved TA signal $I(t)$ at each pixel is Fourier transformed at a chosen phasor frequency $\omega$, yielding two-dimensional phasor coordinates defined by the real and imaginary components:
    \begin{align}
        g(\omega)   &=  \frac{\int I(t)\cos(\omega t)dt}{\int\lvert{I(t)}\rvert dt} \\
        s(\omega)   &=  \frac{\int I(t)\sin(\omega t)dt}{\int\lvert{I(t)}\rvert dt}
    \end{align}
    Each pixel in the image stack maps to a single point in the phasor plane, and species with similar transient absorption  dynamics occupy similar areas in the phasor plane. Adapted phasor analysis has previously been used to distinguish chromophores including hemoglobin, eumelanin, pheomelanin, and surgical ink in pump-probe images \cite{robles2012phasor, robles2015pump}. Here, we extend this framework to two tasks: first, to quantify the heterogeneity of molecular dynamics within and across distinct melanin populations; and second, in combination with machine learning, to assess the separability of different melanin species by their transient absorption signatures. A detailed description of the phasor implementation is provided in the Supplementary material.

    \subsubsection{Classification}
    \label{ch:classification}
    Adapted phasor analysis provides a convenient visualization of transient absorption dynamics by reducing the full TA curve to two scalar coordinates, but is not directly suited to quantifying inter-species separability. To complement phasor visualization, we trained a support vector machine (SVM) classifier on the two-dimensional phasor coordinates $(g(\omega), s(\omega))$ to assess separability. An SVM was chosen because it performs well on classifying multi-exponential decays and TA curves \cite{kastenholz_noninvasive_2024}. Because phasor analysis is a projection of the full TA curve onto two dimensions, we trained a second SVM classifier on the full curve. 
    
    All analyses were performed using scikit-learn \cite{pedregosa2011scikit} and imbalanced-learn \cite{lemaavztre2017imbalanced}. The dataset was partitioned into a held-out test set (20\%) and a training 
    set (80\%), with stratified splitting to preserve class proportions across partitions. Random undersampling was applied within the training set to address class imbalance prior to model fitting. SVM hyperparameters $C$ and $\gamma$ were optimized via 5-fold cross-validation on the training set using balanced accuracy as the optimization criterion.
    
    Because our experimental unit is the ROI rather than the individual pixel, ROI-level predictions were obtained by majority vote: the trained classifier was applied to all pixels within each image stack, and the majority predicted class was assigned as the ROI-level label. This aggregation reduces sensitivity to noise, yielding one prediction per ROI for final performance evaluation on the held-out test set.

\section{Results}
    \subsection{Transient Absorption Curves}
\begin{figure}[ht]
    \centering
    \includegraphics[width = \linewidth]{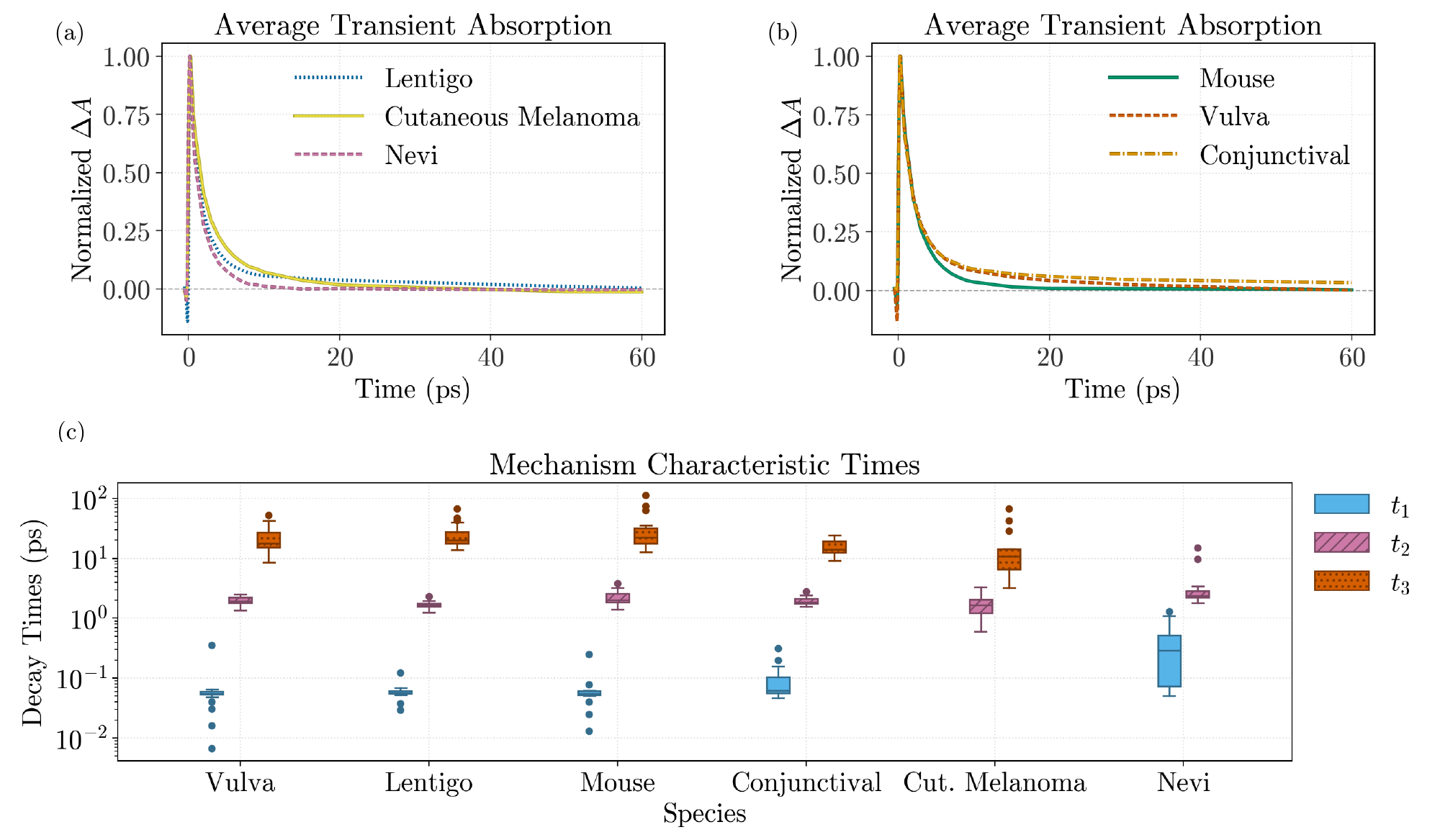}

    \caption{
    Comparison of transient absorption dynamics and extracted characteristic times for the investigated melanin species. 
    (a) Normalized transient absorption traces of melanin from cutaneous malignant melanoma, nevi, and lentigo. (b) Normalized transient absorption traces of mouse, vulva, and conjunctival melanin.  (c) Distribution of fitted characteristic times $t_1$, $t_2$, and $t_3$ shown on a logarithmic scale.}
    \label{fig:ta_and_times}
    \end{figure}

    Figure~\ref{fig:ta_and_times} shows the spatially averaged pump–probe transient absorption (TA) curves for each melanin population. The overall dynamics are qualitatively similar across all six species: a transient gain feature (negative signal) is observed at ultrafast timescales (<150 fs), followed by multiple positive excited-state absorption (ESA) components that decay on picosecond timescales.
    
    The short-lived gain signal (negative) can arise from stimulated emission from an electronically excited state, stimulated Raman scattering (SRS) or other (nearly instantaneous) processes involving virtual energy states. Stimulated emission is unlikely given the low fluorescence quantum yield of melanin \cite{Riesz2007a}. Two-photon absorption is a transient loss process and would manifest itself as a positive signal. Cross-phase modulation would require either a spectral filter or an aperture in the detection to transform the phase modulation into an amplitude modulation. We therefore assign the sub-150 fs negative feature to stimulated Raman scattering (SRS) as the energy difference between pump and probe photons corresponds to a Raman shift of $\Delta \tilde\nu\approx 1650 \text{cm}^{-1}$ matching the C=C stretching mode of melanin \cite{huang_raman_2004, capozzi_raman_2005}. Consistent with this interpretation, the negative feature shows the same temporal behavior across all six melanin populations and is not considered further.

    The remaining positive signal is attributed to excited-state absorption (ESA): pump photons promote electronic population from the ground state to an intermediate excited state, increasing the absorption at the probe wavelength and allowing further transitions to higher states. The ESA components exhibit subtle but reproducible differences in amplitude and lifetime across species, making them the primary source of chemically specific contrast in this experiment. The multi-exponential decay of the ESA signal reflects the structural and chemical disorder intrinsic to melanin. As a heterogeneous biopolymer assembled from DHI, DHICA, and pheomelanin-type oligomeric subunits with varying degrees of conjugation and cross-linking, melanin does not possess a single well-defined excited electronic state. Instead, its photophysics are governed by a distribution of chromophoric environments, each contributing differently to the observed relaxation dynamics \cite{tran_chemical_2006, grieco_probing_2020, kohl_ultrafast_2020}. The tri-exponential decay model used here is therefore phenomenological and the three recovered time constants ($t_1, t_2, t_3$) represent effective decay rates averaging over this distribution, rather than discrete electronic transitions.
    
    Average TA curves per region of interest (ROI) were fitted using a tri-exponential decay model as described in chapter \ref{ch:fitting}. For the majority of ROIs, three characteristic lifetimes were recovered and in some cases  fitting returned fewer than three resolvable components, which was attributed to low signal-to-noise ratio or insufficient sampling time, see Supplementary material. Summary statistics for each species are shown in figure \ref{fig:ta_and_times}c. The lifetimes span three timescales. The fastest component $t_1 \lesssim 1$ ps falls below or near the temporal resolution of the instrument ($\approx 150$ fs) for most species, except for nevi. The intermediate component ($1 \text{ps} \lesssim t_2 \lesssim 5$ps) shows inter-species variation comparable in magnitude to intra-species variation, suggesting limited discriminating power. The slowest component ($t_3 \gtrsim 5$ ps) shows the clearest inter-species differences: conjunctival and cutaneous melanoma exhibit systematically shorter $t_3$ values compared to vulvar melanin, lentigo, and the mouse model. Outlying values (circles) in figure \ref{fig:ta_and_times}c are attributed to poor convergence of the tri-exponential fit in ROIs with low melanin content and consequently low signal-to-noise ratio. The overall similarity of TA curves across populations, see figures \ref{fig:ta_and_times}a-b, is consistent with the shared core architecture of melanin across anatomical sites. The population-dependent differences in ESA lifetime and amplitude, while subtle, are reproducible and form the basis for the classification analysis presented in \ref{ch: class}.
    
    \subsection{Melanin Heterogeneity}
    Having established the spatially averaged transient absorption (TA) dynamics of different melanin populations in the previous chapter, we now discuss spatial heterogeneity, i.e. how TA signals vary across individual tumor sections and between populations. We use adapted phasor analysis, see chapter \ref{ch:phasor}, to visualize and quantify signal heterogeneity across melanin populations. In phasor space, pixels with similar TA decay dynamics cluster together, so that a compact distribution reflects high signal homogeneity while a dispersed distribution indicates chemical or structural heterogeneity in the underlying excited state relaxation. 

\begin{figure}[h!]
        \centering
        \includegraphics[width = \linewidth]{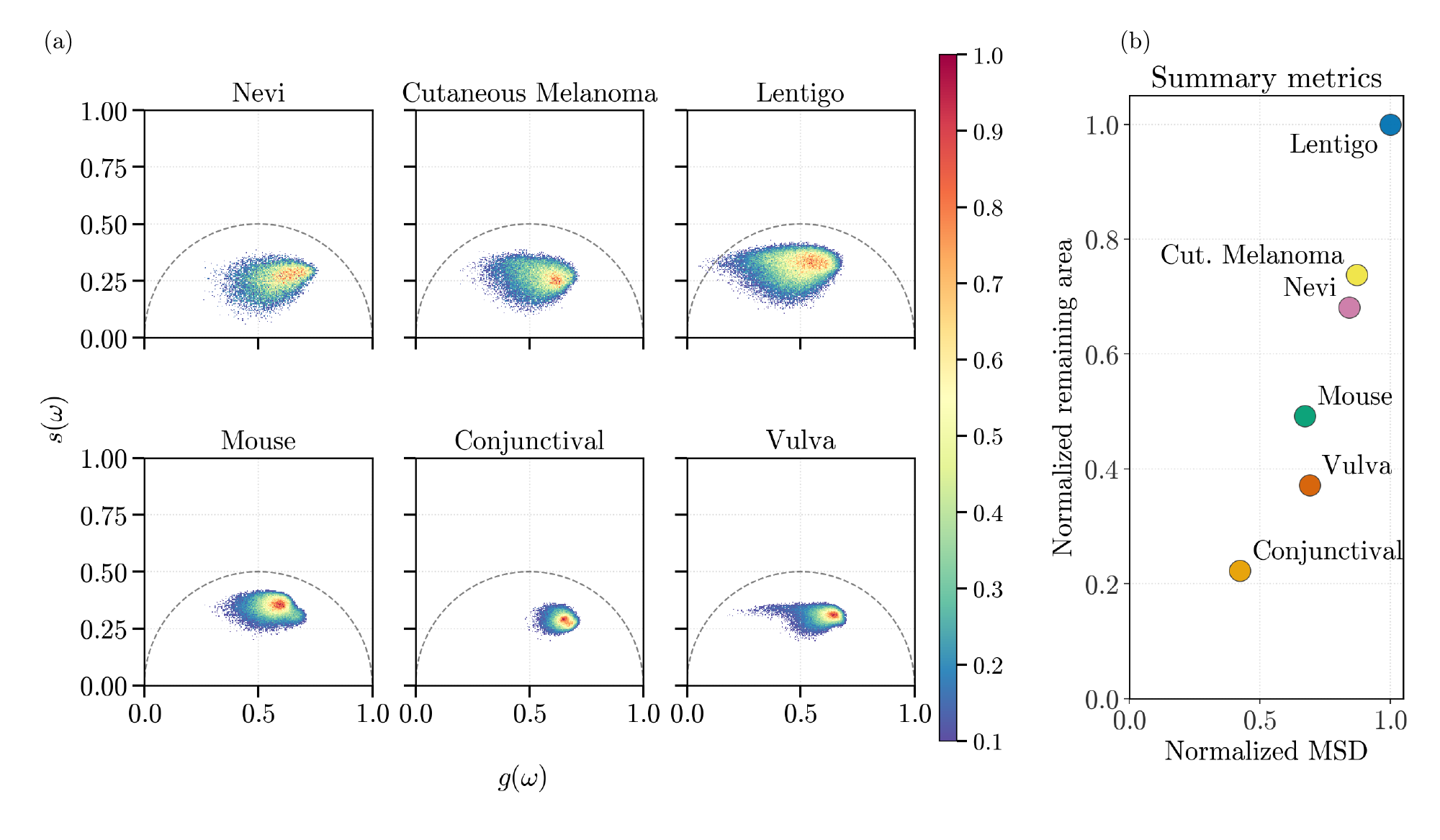}
        \caption{Melanin phasor distributions and summary metrics at $\omega/{2\pi} = 0.25$ THz.}
        \label{fig:phasor_msd_combined}
    \end{figure}

    We selected a phasor frequency of $\omega/2\pi = 0.25$~THz, corresponding to an oscillation period of 4 ps, highlighting ESA differences of similar lifetimes where the dynamics are most pronounced (see figure \ref{fig:ta_and_times}). Cumulative phasor plots for all six melanin populations at this frequency are shown in figure \ref{fig:phasor_msd_combined}a. 
    
    While the distributions overlap and occupy a broadly similar region of phasor space, their centers, extents, and shapes differ markedly. To quantify heterogeneity beyond visual inspection, we introduce two metrics: First, the mean square displacement (MSD) of phasor points relative to the species mean phasor coordinate. Second, the phasor area, defined as the geometric spread of the phasor distribution. MSD weights contributions by density making it sensitive primarily to the spread of the dominant signal population and robust to low-SNR outliers while phasor area captures the total spread regardless of density. Both metrics are shown in figure \ref{fig:phasor_msd_combined}b.    
   Melanin from conjunctival lesions yields the lowest values on both metrics, indicating high signal homogeneity consistent with its relatively uniform eumelanin composition. Lentigo reaches the maximum on both MSD and area, identifying it as the most heterogeneous class in this dataset. This is consistent with the variable pigmentation seen across lentigo lesions. Cutaneous melanoma and nevi score similarly high on both metrics, reflecting the well-documented intratumoral heterogeneity of melanin composition in human cutaneous melanoma and premalignant lesions, while mouse melanoma and vulvar melanin occupy intermediate positions. The concordance between MSD and area rankings is consistent with observed differences reflecting genuine population-level heterogeneity.
   
\begin{figure}[h!]
        \centering
        \includegraphics[width= \textwidth]{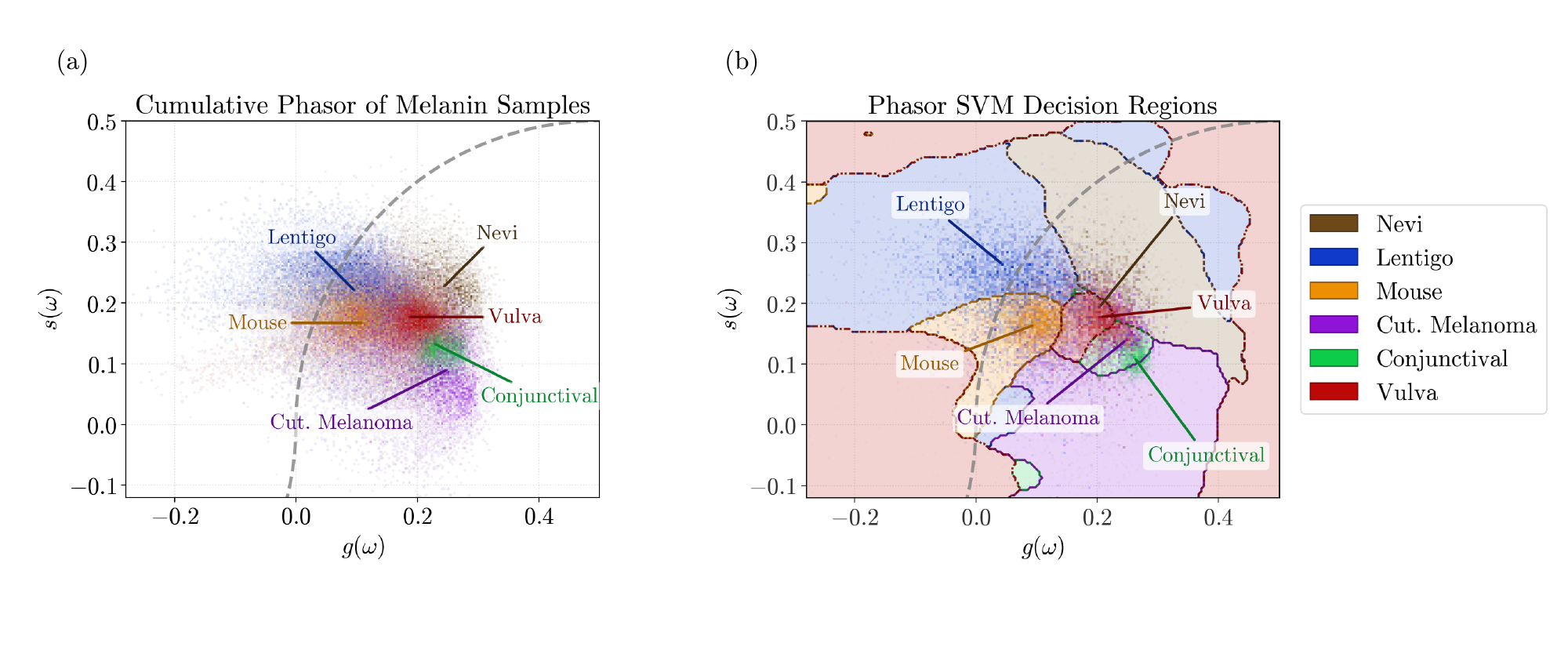}
        \caption{(a) Phasor plot of $\text{n} = 109$ melanin ROIs with a phasor frequency of $\omega/{2\pi} = 0.96$ THz. Labels of melanin species point towards the phasor pixel with the highest density. (b) Plot of phasor pixels used in SVM validation with corresponding decision boundaries.}
        \label{fig:bigphasor}
    \end{figure}

    Figure \ref{fig:bigphasor} shows a cumulative phasor plot of all n = 109 ROIs color-coded by population, computed at a phasor frequency of $\omega/2\pi = 0.96$~THz. This higher frequency was chosen to optimize inter-population separability: faster phasor frequencies emphasize shorter decay components where species-dependent differences in $t_1$ and $t_2$ contribute more strongly to the phasor coordinates. At this frequency, some melanin populations, e.g. from nevi and from lentigo, appear largely separable, while melanin from vulvar and conjunctival lesions show substantial overlap, occupying nearly the same region of phasor space. This overlap is consistent with the similarity of their $t_2$ and $t_3$ distributions (see figure \ref{fig:ta_and_times}). This distribution of populations across phasor space highlights that pump–probe microscopy combined with phasor analysis provides a viable framework for distinguishing melanin from different anatomical sites and tumor origins.
    
    \subsection{Melanin Classification}
    \label{ch: class}
    We trained two pixel level SVM classifiers, see chapter~\ref{ch:classification}, one based on the phasor coordinates shown in figure~\ref{fig:bigphasor}a and one on the full TA curves. The decision boundaries learned by the phasor-based SVM are shown next to the cumulative phasor plot in figure \ref{fig:bigphasor}b. Aggregated by majority vote to the ROI level, the classifier achieves per-class accuracies of 100\% for nevi and lentigo, 67\% for murine melanoma, 50\% for conjunctival melanoma, 33\% for vulvar melanin, and 0\% for malignant cutaneous melanoma. All per‑class accuracies except that of cutaneous melanoma exceed the six-class chance level of $1/6 \approx 0.167$, confirming that the two-dimensional phasor coordinates $(g(\omega), s(\omega))$ encode population-specific information despite their dimensionality reduction. The rank ordering of per-class accuracies is directly consistent with the phasor plot: nevi and lentigo occupy well-separated regions of phasor space and are classified reliably. Conjunctival and vulvar melanins overlap substantially and are consequently more difficult to distinguish. Melanin from melanoma has substantial overlap with melanin from nevi, lentigo, mouse, and conjunctival lesions resulting in no correct classification. Note, however, that on a pixel level basis only around 17\% of malignant melanoma pixels are correctly classified. 
    
    Training the SVM on the full 36-point TA curves instead of the two phasor coordinates substantially improved classification performance, with per-class accuracies reaching 100\% for nevi, lentigo, and murine melanoma. Vulvar and conjunctival melanomas remained  challenging classes at 60\% and 67\%, respectively, consistent with their similar excited-state lifetime distributions (figure~\ref{fig:ta_and_times}). Cutaneous melanoma also had a 67\% classification accuracy. The performance gain of the TA-based over the phasor-based classifier directly reflects the information discarded by the phasor projection: for species whose discrimination relies on subtle differences in their excited state dynamics, the two-coordinate phasor representation is insufficient. Given the small number of ROIs per class in the held-out test set, per-class accuracies should be interpreted as indicative rather than definitive. Complete accuracy tables for all four classifiers (pixel- and ROI-level, phasor- and TA-based) are provided in the Supplementary materials.

\section{Discussion}
Despite sharing a common neural crest origin, melanocytes at different anatomical sites diverge substantially in their local microenvironments, hormonal regulation, and UVR exposure history \cite{cichorek2013skin, mort2015melanocyte}. These biological differences are reflected in the transient absorption dynamics revealed by PPM. Our results indicate that the dominant cause of variation in melanin excited-state dynamics is anatomical site rather than disease state. 

The most striking finding from the phasor heterogeneity analysis is that phasor spread separates cutaneous from non-cutaneous melanin. Melanin from nevi, lentigines, and melanoma arising in chronically UV-exposed skin exhibit the largest phasor distributions, while melanin from conjunctival lesions produces the most compact distribution despite direct solar exposure at the conjunctival surface, and vulvar melanin and the murine model occupy intermediate positions. Two features of this ordering are notable. First, it does not track malignancy as benign lentigo exceeds cutaneous malignant melanoma on both metrics of spread, consistent with the chronically enhanced and irregular melanogenesis of these lesions \cite{matthews_vivo_2011, cardinali2012mechanisms}. Second, it does not follow UVR exposure in a simple dose-dependent manner, since conjunctival melanin is the most homogeneous population in the dataset despite having direct solar exposure. Phasor spread therefore appears to correlate more strongly with the tissue environment in which the melanin is produced and stored, rather than the disease state or the cumulative irradiation. 

In cutaneous tissue, UVR drives a biphasic response: immediate pigment darkening through photo-oxidation of pre-existing melanin and redistribution of melanosomes from a perinuclear to peripheral dendritic location, followed by delayed tanning involving increased melanocyte activity, dendricity, and altered melanosome synthesis and packaging \cite{brenner2008protective, garmyn_mechanisms_2018}. Autocrine mediators including prostaglandins $\text{E}_2$ and $\text{F}_{2\alpha}$ further amplify this response by increasing melanocyte dendricity and facilitating melanosome transfer to keratinocytes \cite{brenner2008protective}. Moreover, UVR alters melanin's  eumelanin/pheomelanin content \cite{hennessy_eumelanin_2005, brenner2008protective}, aggregate size and chromophore redox state \cite{gallas_structural_2000, pezzella_disentangling_2009}, parameters that Simpson et al. demonstrated to produce measurable changes in near-infrared transient absorption dynamics in controlled experiments \cite{Simpson2014}. Because phasor spread does not order the populations by malignancy or irradiation, it should be interpreted alongside morphological features rather than as a standalone indicator of disease severity. 

The compact phasor distribution of conjunctival melanin is consistent with the stability of melanin from conjunctival lesions, similar to observations by Wilson et al. \cite{wilson_imaging_2013}. Unlike uveal and retinal pigment, conjunctival melanocytes lie at the ocular surface and receive direct solar exposure, and conjunctival melanomas carry the UVR-associated mutational signature SBS7 \cite{mundra2021ultraviolet}. The compact distribution therefore cannot be attributed to a lack of exposure, and rather indicates that UVR incident on melanin is not sufficient to generate the heterogeneity seen in cutaneous lesions. While both cutaneous and conjunctival lesions receive direct UVR exposure, the conjunctiva is made of stratified non-keratinized epithelium lacking rete ridges, where the papillary and reticular organization of cutaneous stroma do not apply \cite{whittington2024melanocytic}. 

The intermediate spread of vulvar melanin, drawn from the cohort of Robles et al. \cite{Robles2017}, likely reflects the mixture of lesion subtypes within that cohort including homogeneous atypical genital nevi as well as heterogeneous invasive melanomas contributing spread, rather than photochemical changes \cite{wohlmuth2021vulvar}. The murine BRAF\textsuperscript{V600E} model is distinct from all human populations in phasor space, consistent with its single-mutation, dermally located, pharmacologically induced melanogenesis that bypasses the complex extrinsic influences including UVR, paracrine signaling from surrounding tissue, and immune interactions that collectively shape melanin composition in human cutaneous tumors \cite{goel2009melanocytic, larue2012origin}. 

The excited-state lifetimes are broadly consistent with multi-component relaxation in melanin: the sub-picosecond component reflects ultrafast charge-transfer polaron pair recombination \cite{kohl_ultrafast_2020}, while the picosecond components reflect the distribution of chromophore conjugation lengths underlying melanin's broadband absorption \cite{tran_chemical_2006, Ju2018}. 

The SVM classifier (based on full pixel-level TA curves) achieves a balanced test accuracy of 0.5788, well above the six-class chance level of $1/6 \approx 0.167$, confirming that transient absorption dynamics encode sufficient information for statistically meaningful melanin subtype discrimination. Low per-class accuracies for vulvar (0.46) and conjunctival (0.53) are consistent with their overlap in phasor space and similar lifetime distributions. Meanwhile, lentigo (0.64), cutaneous melanoma (0.65), and nevi (0.74) are classified most accurately, reflecting their photophysical distinctiveness.

These findings position PPM as a platform for optical melanin phenotyping across anatomical sites, complementing its established role in within-tissue lesion stratification \cite{robles2015pump, matthews2011pump, wilson_imaging_2013, Robles2017}. The ability to resolve tissue-dependent chemical heterogeneity as a quantifiable phasor metric opens new avenues for understanding melanoma pathogenesis and evaluating pigmented lesions at anatomically challenging sites.

\section{Conclusion} 
We applied pump-probe microscopy to systematically compare melanin from six biologically and anatomically distinct melanocytic populations across 21 samples and 109 ROIs. Transient absorption dynamics contained a multi-exponential excited-state absorption across anatomical site. Species-dependent differences emerged in the slowest ESA component and in the spread of pixel-level dynamics: phasor heterogeneity metrics ranked lentigo and cutaneous melanoma highest and conjunctival melanin lowest, separating cutaneous and non-cutaneous lesions rather than malignancy or UVR radiation. Lentigo, a benign lesion, exceeds cutaneous melanoma in heterogeneity, has the largest phasor spread and thus, melanin heterogeneity  should be interpreted alongside morphology rather than as a standalone marker in the context of melanoma disease severity. Conjunctival melanin was the most homogeneous population despite direct solar exposure, indicating that irradiation of pigmented tissue is not sufficient to generate the heterogeneity seen in skin. SVM classification on full TA curves separated most subtypes well above chance, confirming that anatomical origin is encoded in melanin excited-state dynamics. Together, these results extend PPM toward label-free optical melanin phenotyping across tissue compartments, with potential application to lesions at diagnostically challenging sites and to longitudinal monitoring of melanin.

\section*{Funding}
This work was supported by the National Institutes of Health (NIH) under award 1R21-CA280272-01A1 and by the U.S. Department of Defense through the U.S. Army Medical Research Acquisition Activity (USAMRAA) under award ME220082.

\section*{Acknowledgment}
We thank Anna Sheinberg, Xiaotian Feng, Heidi Kastenholz, Angelica Selim, Georgia Beasley, and Loren Smith for their stimulating discussions.

\section*{Disclosures}
The authors declare no conflicts of interest.

\section*{Data availability}
The imaging data supporting this study contain patient-derived specimen information and are therefore not publicly available due to privacy restrictions. De-identified transient absorption datasets, extracted fit parameters, and the analysis and classification code are available from the authors upon reasonable request, subject to institutional data-use agreement and IRB approval.

\section{Supplemental information}
\subsection{Data Acquisition}
Prior to imaging, pump and probe lasers were aligned temporally and spatially. Using a Rhodamine 6G sample, we approximate the cross-correlation of the two pulses because of the instantaneous two-photon absorption mechanism that occurs. From this we extract the individual pulse lengths using a Gaussian fit. Likewise, spatial overlap was verified using 500 nm diameter polystyrene beads that exhibit two-photon fluorescence. Pump and probe beams were individually adjusted until the extracted positions as fit by 3D Gaussians to the beads were within half a standard deviation of each other. 
\subsection{Pre-Processing of Image Stacks}
\label{ch:pre-processing}
Before analyzing the pump-probe stacks, the regions of each image that corresponded to surgical ink were redacted. The images were visually inspected for surgical ink on brightfield images. Furthermore, signal intensity of the surgical ink strongly exceeds that of melanin, allowing for easy identification of ink regions as shown in Figure \ref{fig:heteroTAcurve}c. To improve signal-to-noise ratio, 2x2 pixel were spatially averaged. An optimal Li threshold was calculated for the regions of interest to redact noisy pixels whose projection magnitude were below the calculated threshold amplitude of $0.073$. To eliminate signals with lifetimes much longer that the inter-pulse spacing of 12.5 ns, signals at negative time delays were averaged and subtracted from the entire image stack. Finally, transient absorption curves per pixel were normalized by their maximum amplitude. 

\subsection{Fitting}
\label{ch:supp_fitting}
We used a multi-exponential fit model to fit the averaged transient absorption curves per region of interest following the method presented in \cite{grass2022contrast}. We used a Levenberg-Marquardt algorithm to fit the transient absorption curves using both a bi-exponential and a tri-exponential model, and chose the tri-exponential model as the best fit model as the one with the lowest AIC score. Thus, the pump-probe transient absorption curves are modeled as a superposition of three decay mechanisms, one instantaneous mechanism, and one long lived signal offset. Each decay process is modeled as follows 
$$\text{signal}(t, \tau, a) = \frac{a}{2} \exp\left(-\frac{t}{\tau} + \frac{t_{pu}^2 + t_{pr}^2}{2\tau^2}\right)
\left[
1 + \operatorname{erf}\left(
\frac{t}{\sqrt{2(t_{pu}^2 +t_{pr}^2)}} - \frac{\sqrt{t_{pu}^2 + t_{pr}^2}}{\sqrt{2}\tau}
\right)
\right]$$ where $t_{pu}$ is the pulse width of the pump beam, and $t_{pr}$ is the pulse width of the probe beam. For our setup, $t_{pr}=t_{pu} = 0.1 \text{ ps}$. We model our offset and instantaneous processes as $$\text{offset}(t,a) = \frac{a}{2} \left[1 + \operatorname{erf}\!\left(\frac{t}{\sqrt{2(t_{pu}^2 + t_{pr}^2)}}\right)\right]$$ 
$$\text{instantaneous}(t,a) = \frac{a}{2}  \, \exp\!\left(-\frac{t^2}{2\,(t_{pu}^2+t_{pr}^2)}\right)$$

For 14 regions of interest out of the total 109, the tri-exponential fit would fail to converge. In particular in the cases of cutaneous melanoma and nevi, which made up a majority of the ill-fitting exceptions, a bi-exponential model consisting of two decays, one instantaneous mechanism, and one offset was used instead. The fitted lifetimes were sorted into three regimes from shortest to longest: $t_1$ ($< 1$ ps), $t_2$ ($1-5$ ps), and $t_3$ ($> 5$ ps). For the bi-exponential fits, the times were sorted into the same categories. Additionally, for 2 regions of interest, an interplay between the longest lifetimes and the constant offset was observed, i.e. the exponential decay and the constant offset counteract each other to obtain an approximately linear decrease in signal. Consequently, the longest extracted time constants were discarded as they are not on the measurement timescale. Moreover, 1 lentigo region of interest failed to fit for both the tri-exponential and bi-exponential models such that a mono-exponential was used instead. 

\subsection{Adapted Phasor Analysis}
\label{ch:supp_phasor}
Adapted phasor analysis has been previously used to distinguish chromophores such as hemoglobin, eumelanin, pheomelanin, and surgical ink \cite{robles2012phasor, robles2015pump}. Here, we use this approach to inspect the differences in heterogeneity and molecular dynamics across distinct melanin populations. In practice, the transient absorption signal $I(t)$ at each pixel is transformed into phasor coordinates $g(\omega) \text{ and } s(\omega)$ components, corresponding to the real and imaginary parts of the Fourier transform at a specific phasor frequency $\omega$. These components are defined as: 
\begin{equation}
    g(\omega)=\frac{\int I(t)cos(\omega t)dt}{\int\lvert{I(t)}\rvert dt}
\end{equation}
\begin{equation}
    s(\omega)=\frac{\int I(t)sin(\omega t)dt}{\int\lvert{I(t)}\rvert dt}
\end{equation}

\begin{figure}[ht!]
    \centering
    \includegraphics[width=0.95\linewidth]{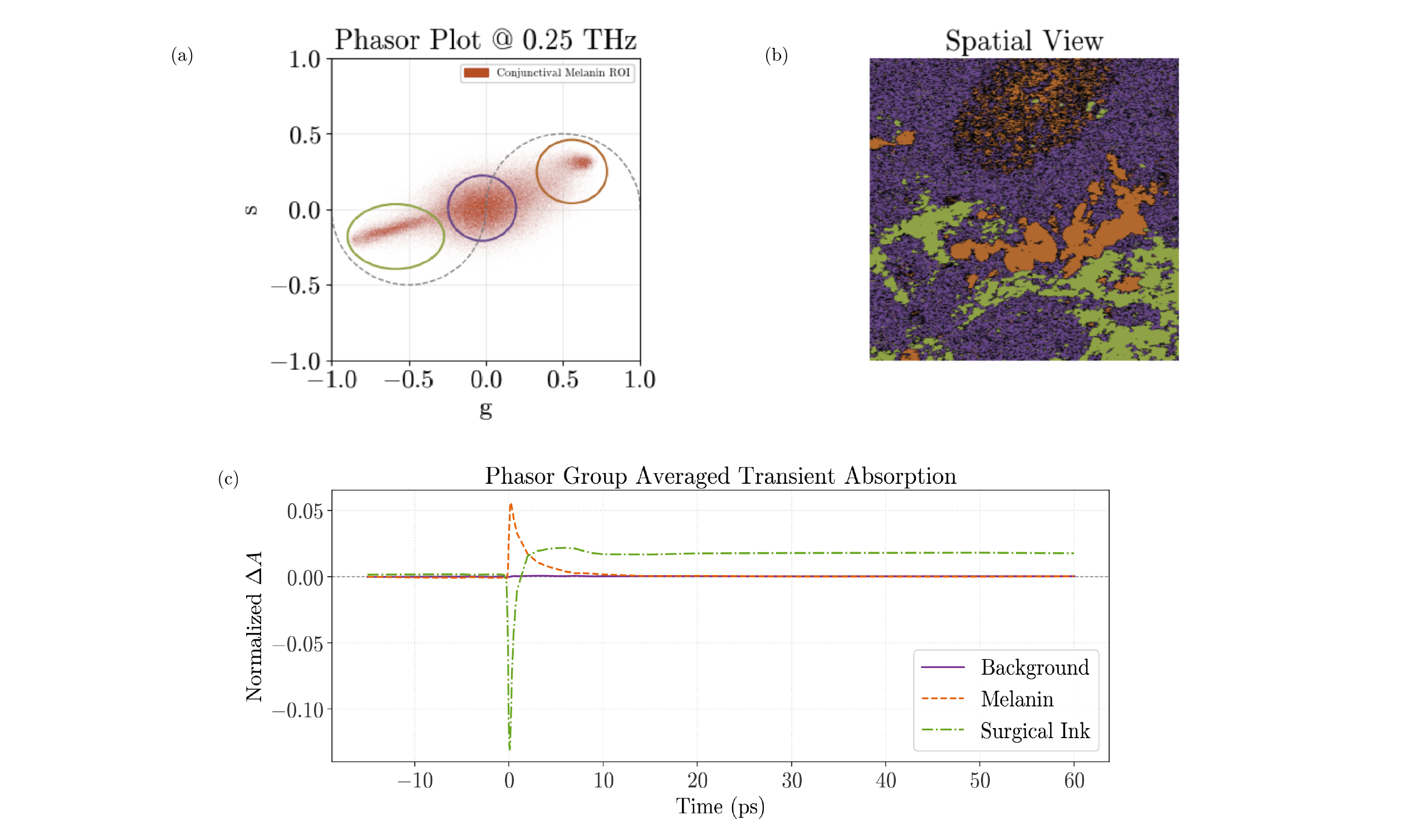}
    \caption{(a) Phasor plot of a conjunctival region of interest with circled clusters (b) Spatial view of the pixels on the PPM image with corresponding circled phasor values (c) Averaged transient absorption of circled phasor values}
    \label{fig:heteroTAcurve}
\end{figure}

Therefore, each pixel in a pump probe image stack is mapped  to a point in phasor space $(g(\omega), s(\omega))$, see for example Figure \ref{fig:heteroTAcurve}a. In the main manuscript, we applied a threshold of $\alpha = 0.1$ to suppress extraneous bins of the phasor points on the plot that had a density of less than $10\%$ of the maximum density of the corresponding phasor species.

We can use phasor plots to visualize heterogeneity of transient absorption dynamics. For example, Figure \ref{fig:heteroTAcurve} shows a phasor plot of conjunctival melanoma  prior to  preprocessing. We identified separate clusters in the phasor plot. Looking at their associated average transient absorption curves in Figure \ref{fig:heteroTAcurve}c, we can assign species labels associated with melanin, surgical ink, and background. For example, the red circled region on the rightmost area of the plot corresponds to pixels that resemble melanin, which is dominated largely by positive signal. In contrast the green circled phasor region on the leftmost area of the plot corresponds to pixels that resemble surgical ink, which is dominated largely by an intense negative signal. The circled purple region in the center shows largely signal background. These differing dynamics are represented by the spatial spread of the phasor histogram.  Heterogeneity was quantified using metrics derived from the spatial distribution of phasors. The radial standard deviations $\sigma_r$ of each melanin subtype cumulative phasor and the number of surviving bins after thresholding was implemented. While $\omega/{2\pi} = 0.96$ THz appeared to give the best visual separation between different subtypes, $\omega/{2\pi} = 0.25$ THz appeared to give the most conclusive measures for separation based on levels of spread.

\subsection{Classification}
\label{ch:supp_classification}
We trained a support vector machine (SVM) to classify the melanin transient absorption curves for each melanin subtype. A SVM works to classify data by finding a decision boundary that maximizes the margins between classes.  We used scikit-learn 1.8 and imbalanced-learn 0.14.1 for training, validating, and testing \cite{pedregosa2011scikit, lemaavztre2017imbalanced}. We split the data into test/train and a holdout validation set with a 4:1 ratio using scikit-learn StratifiedGroupKFold to ensure that individual regions of interest did not appear in both sets. The test/train portion was further split using StratifiedGroupKFold via a 5-fold stratified grouped cross validation during hyperparameter optimization. Within each fold, class imbalance was addressed using imblearn RandomUnderSampler, which randomly undersamples the majority class to match minority class distributions. All features of the TA curves were standardized using StandardScaler. Hyperparameter optimization was performed using scikit-learn GridSearchCV over the SVM parameters $C$ and $\gamma$, with balanced accuracy as the scoring metric. The optimal classifier was selected based on the highest mean cross-validation balanced accuracy across folds. The final model is then evaluated on the hold out validation set. Because the validation set retained the original class imbalance, classifier performance was assessed using balanced accuracy, which is the average recall. 
\begin{table}[h!]
    \centering
    \begin{tabular}{c|cccccc}
         Classifier &   Nevi   & Lentigo   &  Malignant Melanoma  & Vulva     & Conjunctival    & Mouse \\  
         \toprule
         phasor, pixel-level  & 0.41  & 0.54  & 0.17  & 0.26  & 0.30  & 0.52 \\
         phasor, ROI-level    & 1.00  & 1.00  & 0  & 0.33  & 0.50  & 0.67 \\
         \hline
         TA, pixel-level      & 0.74  & 0.64  & 0.65  & 0.46  & 0.53  & 0.45 \\
         TA, ROI-level        & 1.00  & 1.00  & 0.67  & 0.60  & 0.67  & 1.00 \\
        \bottomrule
    \end{tabular}
    \caption{Validation accuracies of SVM classifiers on phasor-based pixel level, phasor-based ROI level, TA-based pixel level, and TA-based ROI level.}
    \label{tab:placeholder}
\end{table}

\bibliographystyle{unsrt}
\bibliography{sample}

\end{document}